\documentclass[12pt]{article}
\usepackage{graphicx}

\begin{document}

\title{\bf Isentropic perturbations of a chaotic domain}

\author{I.~I.~Shevchenko\/\thanks{E-mail:~iis@gao.spb.ru} \\
Pulkovo Observatory of the Russian Academy of Sciences \\
Pulkovskoje ave.~65/1, St.Petersburg 196140, Russia}
\date{}

\maketitle


\begin{center}
Abstract
\end{center}

\noindent Three major properties of the chaotic dynamics of the
standard map, namely, the measure $\mu$ of the main connected
chaotic domain, the maximum Lyapunov exponent $L$ of the motion in
this domain, and the dynamical entropy $h = \mu L$ are studied as
functions of the stochasticity parameter $K$. The perturbations of
the domain due to emergence and disintegration of islands of
stability, upon small variations of $K$, are considered in
particular. By means of extensive numerical experiments, it is
shown that these perturbations are isentropic (at least
approximately). In other words, the dynamical entropy does not
fluctuate, while local jumps in $\mu$ and $L$ are significant.

\bigskip


\newpage

The standard map is an important object of studies in nonlinear
dynamics, mainly because it describes local behaviour of the
separatrix map, in a wide range of values of
parameters~\cite{C78,C79}. The separatrix map represents the
motion in the vicinity of the separatrices of a nonlinear
resonance subject to periodic
perturbation~\cite{C78}--\cite{LL92}. General properties of the
standard map, often in interrelation with those of the separatrix
map, were considered and studied in detail
in~\cite{C78}--\cite{GL00} and many other works. Apart from its
relation to the separatrix map, the standard map serves as an
important independent mechanical and physical
paradigm~\cite{C96,M92}.

Nonlinear resonances are ubiquitous in the problems of modern
mechanics and physics. Under general conditions
(see~\cite{C77,C79,LL92}), the model of a nonlinear resonance is
provided by the nonlinear pendulum with periodic perturbations.
Chirikov~\cite{C77,C79} derived the separatrix map describing the
motion in the vicinity of the separatrices of a nonlinear
resonance in the perturbed pendulum model. Convenient equations
for the map, used e.~g.\ in~\cite{CS84,S98a}, are the following:

\begin{eqnarray}
     y_{i+1} &=& y_i + \sin x_i, \nonumber \\
     x_{i+1} &=& x_i - \lambda \ln \vert y_{i+1} \vert + c
                   \ \ \ (\mbox{mod } 2 \pi),
\label{sm1}
\end{eqnarray}

\noindent where $x$, $y$ are dynamical variables, $\lambda$ and
$c$ are constant parameters.

The standard map is the linearization of the separatrix map in the
action-like variable $y$ near a fixed point. The map is given by
the equations

\begin{eqnarray}
     y_{i+1} &=& y_i + K \sin x_i \ \ \ (\mbox{mod } 2 \pi), \nonumber \\
     x_{i+1} &=& x_i + y_{i+1} \ \ \ (\mbox{mod } 2 \pi),
\label{stm}
\end{eqnarray}

\noindent where $K$ is the so-called stochasticity
parameter~\cite{C78,C79}. On the other hand, the motion near the
separatrices of the resonances of the standard map is described by
the separatrix map. Therefore major properties of the separatrix
map are inherently determined by the properties of the standard
map, and vice versa.

In such a way basic properties of the standard map determine the
chaotic behaviour near separatrices of a nonlinear resonance. To
elucidate these properties is the major goal of the present study.

The calculation of the Lyapunov characteristic exponents (LCEs) is
one of the most important tools in the study of the chaotic
motion. The LCEs characterize the rate of divergence of
trajectories close to each other in phase space. A nonzero LCE
indicates chaotic character of motion, while the maximum LCE equal
to zero is a signature of regular (periodic or quasi-periodic)
motion. The quantity reciprocal to the maximum LCE characterizes
the predictability time of the motion.

Let us consider two trajectories close to each other in phase
space. One of them we shall refer to as {\it guiding} and the
other as {\it shadow}. Let $d(t_0)$ be the length of the
displacement vector directed from the guiding trajectory to the
shadow one at an initial moment~$t = t_0$. The LCE is defined by
the formula~\cite{LL92}:

$$L=\limsup_{{t \to \infty} \atop {d(t_0) \to 0}}
         {1 \over {t-t_0}} \ln{d(t) \over d(t_0)} \, . $$

\noindent The LCEs are closely related to the dynamical
entropy~\cite{P76,BGS76}. For the Hamiltonian systems with 3/2 and
2 degrees of freedom, Benettin {\it et al.} proposed the relation
$h = \mu L $~\cite[Eq.~(6)]{BGS76}, where $h$ is the dynamical
entropy, $\mu$ is the relative measure of the connected chaotic
domain where the motion takes place, $L$ is the maximum LCE of the
motion. This formula is approximate. Benettin {\it et
al.}~\cite{BGS76} applied it in a study of the chaotic motion of
the H\'enon--Heiles system.

In what follows, we present numerical data on the chaotic domain
measure $\mu$, the maximum LCE $L$, and their product $h$ for the
standard map. A traditional ``one trajectory method'' (OTM) has
been used for calculation of $\mu$. It consists in computing the
number of cells explored by a single trajectory on a grid exposed
on phase plane. A strict but computationally much more expensive
approach for measuring $\mu$ consists in calculating the values of
the coarse-grained area of the chaotic component for a set of
various resolutions of the grid, in order to find the asymptotic
value of $\mu$ at the infinitely fine resolution
(see~\cite{UF85}). However, a good correspondence of our results
on the measure of small regular islands with independent
theoretical inferences (see below) verifies that the resolutions
we use are already fine enough for the accurate determination of
$\mu$. A ``current LCE segregation method'' (CLSM) has been
employed in some cases for verification of the obtained values of
$\mu$. It is based on an analysis of the differential distribution
of the computed values of the Lyapunov exponents (current LCEs) of
a set of trajectories with the starting values generated randomly
or on a regular grid on phase plane. The OTM and CLSM were both
proposed and used by Chirikov~\cite{C78,C79} in computations of
$\mu$ for the standard map. Analogous methods were used
in~\cite{SM03} in computations of the chaotic domain measure in
the H\'enon--Heiles problem.

Fig.~1 presents our numerical results on $\mu(K)$, $L(K)$, and
$h(K)$ at $K \in [0.5, 1]$. Fig.1a illustrates discontinuity of
the obtained $\mu(K)$ function. The OTM has been used for
constructing this plot; the grid is $2000 \times 2000$ pixels on
phase plane ($x$, $y$) $\in [0, 2 \pi] \times [0, 2 \pi]$. The map
has been iterated $n_{it} = 10^8$ times at each value of $K$; the
step in $K$ is equal to $0.001$.

$L$ is the maximum Lyapunov exponent of the motion in the main
chaotic domain; $h = \mu L$. Each value of $L$ presented in Fig.1b
has been computed simultaneously with $\mu$ for the same
trajectory. Everywhere in this work the presented values of the
maximum LCEs have been computed by the tangent map method
described e.~g.\ in~\cite{C78,C79}.

As follows from Fig.1b, the dynamical entropy $h = \mu L$ appears
to be sharply different, as a function, from the co-products $\mu$
and $L$: the dependences $\mu(K)$ and $L(K)$ are discontinuous,
while their product $h(K)$, on the contrary, looks continuous and
monotonous.

At moderate values of $K$ (at approximately $K < 4$), the
discontinuities in $\mu(K)$ and $L(K)$ are conditioned by the
process of absorption of minor chaotic domains by the main chaotic
domain, while $K$ increases; at larger values of $K$ (at
approximately $K > 4$), it is conditioned by the process of birth
and disintegration of new regular islands. The most prominent jump
of $\mu(K)$ at $K \approx 0.9716$ is conditioned by the absorption
of the chaotic domain associated with the separatrices of the
half-integer resonance.

The thin solid line in Fig.1b depicts a theoretical approximation
of the $h(K)$ dependence. It is given by the function

\begin{equation}
h(K) = A K^{1/2} \exp \left( -\frac{\pi^2}{K^{1/2}} \right)
\label{hexp}
\end{equation}

\noindent with $A = 929.6 \pm 4.0$; the latter value is obtained
by fitting the observed dependence.

Form~(\ref{hexp}) for the approximating function is derived in the
following way. The maximum mutual divergence of the branches of
the splitted separatrix of the integer resonance of
map~(\ref{stm}) near the point $x = \pi$ is directly proportional
to $K^{-1/2} \exp \left( - \pi^2 K^{-1/2} \right)$; this relation
follows from Eqs.~(1.14) and (1.15) in~\cite{VC98}. The divergence
of the separatrices characterizes the width of the chaotic layer,
and, therefore, the total measure $\mu$ of the chaotic domain.

On the other hand, the maximum LCE in the domain is on the average
directly proportional to $K$ (see Fig.1c). This has a theoretical
explanation. Let us proceed from the relation $L=L_{sm}/T$, where
$L_{sm}$ is the maximum LCE of the separatrix map describing the
motion, and $T$ is the average half-period of phase oscillations
in the chaotic layer. The quantity $L_{sm}$ is given by Eq.~(12)
in~\cite{S02}: $L_{sm} \propto \lambda/(1 + 2 \lambda)$, i.~e.\ it
is practically constant at high values of the separatrix map
parameter $\lambda$. For the standard map, $\lambda = 2 \pi
K^{-1/2}$~\cite{C78,C79}, and therefore $\lambda > 2 \pi$ at $K <
1$. The quantity $T$ is directly proportional to $K^{-1}$ at small
enough values of $K$ (see Eq.~(2.16) in~\cite{C78}, or Eq.~(6.18)
in~\cite{C79}). Hence, $L \propto K$ at small enough values of
$K$.

The product of $\mu \propto K^{-1/2} \exp \left( - \pi^2 K^{-1/2}
\right)$ and $L \propto K$ gives function~(\ref{hexp}). A
paradoxical situation here is that we use formulas for $\mu(K)$
and $L(K)$ that are inherently approximate due to the
discontinuous nature of these two functions, to obtain a formula
for a quantity which might be not discontinuous ($h(K)$), and
which may have a smooth analytical representation. Maybe a future
theory will allow to derive such representation directly, without
appealing to the discontinuous co-products.

In Fig.1c the dependence $L(K)$ is presented with very fine
resolution in $K$ (the step in $K$ is $0.0001$) at $K \in [0.1,
1]$. A larger number of iterations is necessary for saturation of
current LCE values at smaller values of $K$; therefore, at $K <
0.35$ each value of $L$ presented in Fig.1c has been computed for
a single trajectory with $n_{it} = 5 \cdot 10^8$, while at $K \ge
0.35$ the former value $n_{it} = 10^8$ has been adopted. The
saturation at smallest values of $K$ was checked by recomputing
the data at $K \in [0.10, 0.11]$ with $n_{it} = 10^9$.

A self-similar wave-like structure seen in the graph in Fig.1c is
due to the absorption of the chaotic layers of integer marginal
resonances by the main chaotic domain on increasing the
stochasticity parameter $K$; on marginal resonances,
see~\cite{S98a}.

The downward spikes seen in the $L(K)$ dependence in Figs.~1b and
1c represent a manifestation of the so-called ``stickiness
effect'' immanent to the chaotic Hamiltonian dynamics in
conditions of divided phase space~\cite{S98b}: a chaotic
trajectory may stick for a long time to the borders of the chaotic
domain, where the motion is close to regular, and therefore the
local LCEs are small. Since the computation time is always finite,
the stickiness effect, in the case of deep stickings, leads to
underestimation of the LCE values; see discussion in~\cite{S98b}.
The spikes are especially pronounced, due to the high resolution,
in Fig.1c. Spikes due to stickings can be also seen in Figs.~2, 3b
and 4b. It is important to note that these negative jumps in $L$
represent merely statistical phenomena affecting single
trajectories. These jumps are not connected to any perturbations
of the chaotic domain measure, contrary to the positive jumps
discussed below.

In Fig.2, the dependences $L(K)$ and $h(K)$ are presented in a
wider range: $K \in [0, 10]$. The values of $\mu$ and $L$ have
been computed simultaneously for a single trajectory taking
$n_{it} = 10^7$. For computing $\mu$, the OTM has been used with
the grid $1000 \times 1000$ pixels on phase plane ($x$, $y$) $\in
[0, 2 \pi] \times [0, 2 \pi]$. The step in $K$ is $0.01$.

The irregularities present in the $L(K)$ dependence are apparently
smoothed out in the $h(K)$ graph, in the broad range of $K$.

The function

\begin{equation}
h(K) = \ln{K \over 2} \label{logk}
\end{equation}

\noindent is depicted in the same Fig.2 (as well as in Figs.~3b
and 4b below). This is the well-known logarithmic law derived by
Chirikov~\cite{C78,C79} analytically by means of averaging the
largest eigenvalue of the tangent map in assumption that the
relative measure of the regular component is small. The presented
numerical data indicates (see Fig.2; as well as Figs.~3b and 4b
below) that the high-$K$ asymptote of the $h(K)$ dependence is
described by the approximating function

\begin{equation}
h(K) = \ln {K \over 2} + {1 \over K^2}, \label{logk2}
\end{equation}

\noindent instead of Eq.~(\ref{logk}). In other words, the
asymptote contains a power-law component, in addition to the
well-known logarithmic one. The same is valid for the $L(K)$
dependence, if one ignores the small (and local in $K$)
distortions of the function due to the accelerator modes and
periodic solutions of higher orders.

The $\mu(K)$ dependence at large $K$ (at $K$ approximately greater
than $6$) is the horizontal line $\mu = 1$ with periodic sequences
of narrow minima of small and decreasing depth. The most prominent
of these sequences correspond to the accelerator modes (emerging
at $K \approx 2 \pi m$; $m=1, 2, \ldots$) and the 4-periodic
solutions (emerging at $K \approx 2 \pi \left(m + \frac{1}{2}
\right)$; $m=1, 2, \ldots$). The less pronounced sequences of
minima correspond to periodic solutions of higher orders.

The local minima in $\mu(K)$ correspond to local maxima in $L(K)$.
The discontinuous patterns in $L(K)$ are removed by the procedure
of multiplication of $L$ by $\mu$. Figs.~3a and 3b illustrate the
removal of a pattern (a jump) in $L(K)$ at a high value of $K$ by
means of plotting of $h = \mu L$ instead of $L$. The pattern is
conditioned by birth and decay of the islands due to a 4-periodic
solution. Figs.~4a and 4b show $\mu(K)$, $L(K)$, and $h(K)$ for a
perturbation due to an accelerator mode.

The OTM has been used for computation of $\mu(K)$ in both Figs.~3a
and 4a; the grid is $2000 \times 2000$ pixels, $n_{it} = 10^8$.
Each value of $L(K)$ has been computed for a single trajectory
simultaneously with $\mu(K)$. The step in $K$ is $\Delta K =
0.001$.

The minimum values of $\mu$ for the both cases of the accelerator
mode and the 4-periodic solution (the minima seen in Figs.~3a and
4a) are in a good agreement with the semi-analytical scaling
$\mu_{reg} = 0.38 K_m^{-2}$ derived by Chirikov~\cite{C96} (see
also~\cite{C78,C79}) for the maximum area of the accelerator mode
islands; here $K_m$ are the values of $K$ at which the maximum
values of area of the regular islands are achieved. The difference
is within $(1 \div 2) \cdot 10^{-4}$ in the both cases. This
agreement verifies good accuracy of our measurements of $\mu$.

Note that the inversely quadratic decay of area of the regular
islands emerging at $K \approx \pi m$; $m=2, 3, \ldots$, i.~e.\
for the both cases of the accelerator modes and the 4-periodic
solutions, was recently rigorously derived by Giorgilli and
Lazutkin~\cite{GL00}. They refer to an unpublished work by
Lazutkin, Petrova, and Svanidze for the numerical confirmation of
this theoretical result.

In the both cases of Figs.~3 and 4, the patterns in $L(K)$ are
substantially eliminated by considering $h$ instead of $L$. What
is the mechanism for the ``maximum LCE---chaos measure'' local
anticorrelations?

Let us consider an island's decay. The small self-similar
variations of $\mu(K)$ and $L(K)$ during the decay are due to the
absorption (by the main chaotic domain) of the chaotic layers of
the chains of islands separating from the main island's border on
increasing $K$. We mark the properties of the whole chaotic domain
just before and after an act of the absorption by the subscripts
$b$ and $a$ respectively. Assume that the absorbed layer has
almost zero local value of the maximum LCE. Then, the global value
of the maximum LCE after an act of the absorption is the value
averaged over the whole accessible area: $L_a = (L_b \mu_b + 0
\cdot \delta \mu)/(\mu_b + \delta \mu)$, where $\delta \mu$ is the
measure of the layer absorbed, $\mu_b$ and $L_b$ are the value of
the measure of the main connected chaotic domain and the value of
the maximum LCE of the motion in this domain before the act of the
absorption. Then, the new value of the dynamical entropy $h_a =
\mu_a L_a = (\mu_b + \delta \mu) \left( \mu_b L_b/(\mu_b + \delta
\mu) \right) = \mu_b L_b$ remains equal to the old one $h_b$,
i.~e.\ $h(K)$ is insensitive to such absorptions and is subject
solely to ``secular'' variation.

This reasoning explains also the smoothing out of the $h(K)$
dependence at $K < 1$, since the growth of the main chaotic domain
at these values of $K$ is conditioned just by the absorption of
the external chaotic layers.

However, this reasoning is not completely satisfactory in the both
cases, because whatever small the local Lyapunov exponent in the
absorbed layer might be, it is not strictly zero. Besides, Fig.~3a
clearly demonstrates that the process of the consequent
absorptions of the layers takes place during the decay of the
pattern, but not during its growth: contrary to a fractal (due to
the absorptions) structure of the $\mu(K)$ dependence during the
decay, this dependence is apparently smooth, of course up to the
accuracy of our graph, during the growth. So, the approximate
conservation of the dynamical entropy is probably not due to the
just described mechanism at intervals of $K$ corresponding to the
growth of islands. In fact, the largest visible deviations (though
small) from the conservation occur at the intervals of growth. Due
to the smallness of these deviations, it is not clear currently
whether they are real, or the adopted resolution of our numerical
experiments is insufficient.

Apart from the need for a complete theoretical explanation in the
considered case of the standard map, an important problem is
whether the perturbations due to the emergence and disintegration
of islands of stability in chaotic domains are isentropic, at
least approximately, in any Hamiltonian system. A hint for an
affirmative answer at least in case of a particular system with
two degrees of freedom can be found in~\cite{SM03}: indeed, the
energy dependence of the dynamical entropy of the chaotic domain
in phase space of the H{\'e}non--Heiles system (Fig.8 of that
paper) is more regular, or smoothed out, in comparison with the
corresponding energy dependence of the maximum LCE (Fig.~2 of that
paper).

Of course, if the system can be approximated locally by the
standard map, the perturbations of the local chaotic behaviour,
according to our numerical results, should be approximately
isentropic.

Let us summarize the main results. The measure $\mu$ of the main
connected chaotic domain of the standard map, the maximum Lyapunov
exponent $L$ of the motion in this domain, and the dynamical
entropy $h = \mu L$ have been studied as functions of the
stochasticity parameter $K$ by means of extensive numerical
experiments. It has been found that the $h(K)$ dependence behaves
like a continuous and monotonous function (up to the experimental
accuracy, of course), while the $\mu(K)$ and $L(K)$ functions are
discontinuous. A semi-analytical approximating relation for the
dynamical entropy at $K < 1$ has been derived.

It has been shown that the process (upon small variations of $K$)
of birth and disintegration of the islands of stability inside the
chaotic domain does not result in fluctuations of the dynamical
entropy $h(K)$ (while $\mu(K)$ and $L(K)$ fluctuate); i.~e.\ these
perturbations are isentropic, at least approximately. A tentative
explanation of the isentropic behaviour has been given.

The author is thankful to B.\,V.~Chirikov for valuable
discussions. This work was supported by the Russian Foundation for
Basic Research (project number 03-02-17356).

\newpage

\begin{figure}
\begin{center}
\begin{tabular}{ll}
a)~\includegraphics[width=6.55cm]{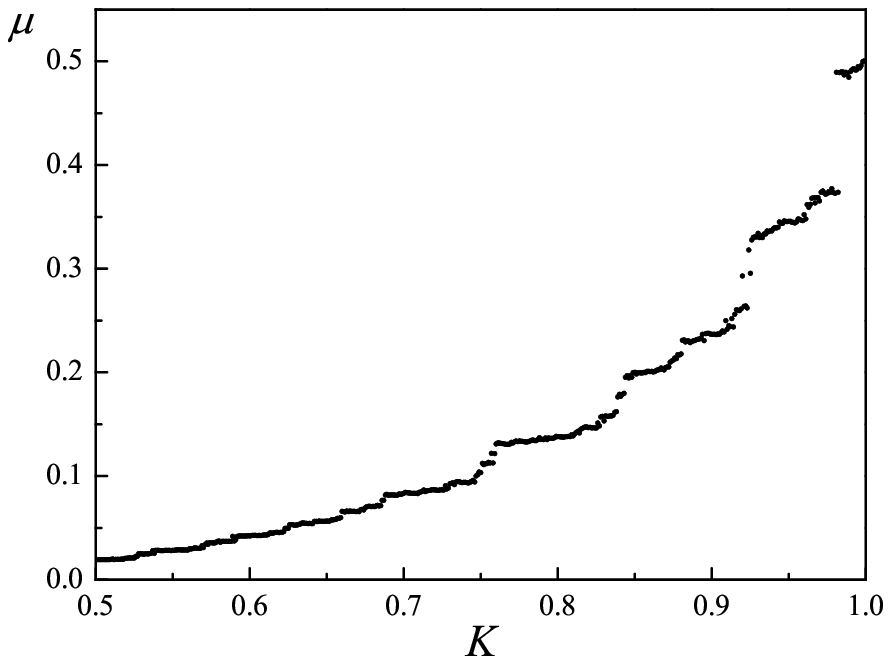} &
b)~\includegraphics[width=7cm]{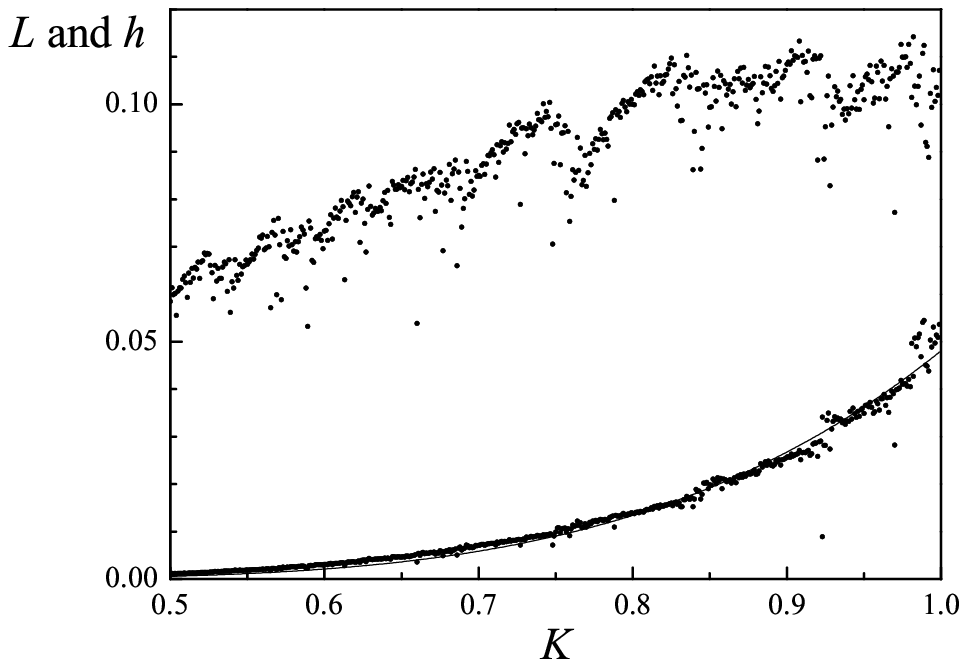}
\end{tabular}
\end{center}
\centering
c)~\includegraphics[width=6.55cm]{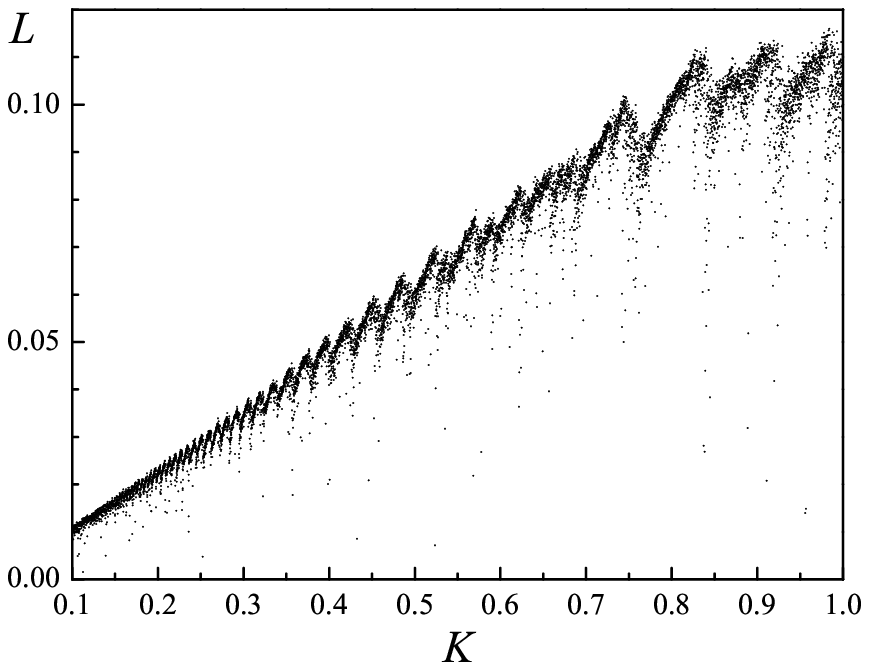}
\caption{The $\mu(K)$ dependence at $K < 1$~(a); $L(K)$, $h(K) =
\mu(K) L(K)$, and the theoretical curve for $h(K)$ given by
Eq.~(\ref{hexp}) (the thin solid line)~(b); $L(K)$ in a higher
resolution~(c).}
\label{fig1}
\end{figure}

\begin{figure}
\centering
\includegraphics[width=7cm]{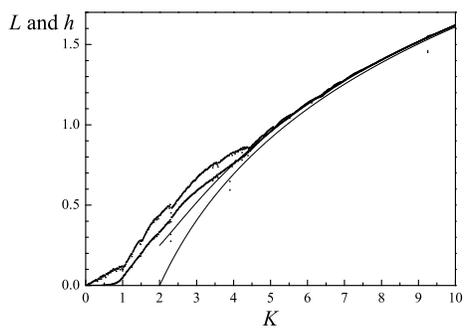}
\caption{$L(K)$ (the upper curve) and $h(K)$ in a wider range of
$K$. The thin solid lines are given by Eqs.~(\ref{logk},
\ref{logk2}).}
\label{fig2}
\end{figure}

\begin{figure}
\begin{center}
\begin{tabular}{ll}
a)~\includegraphics[width=6.8cm]{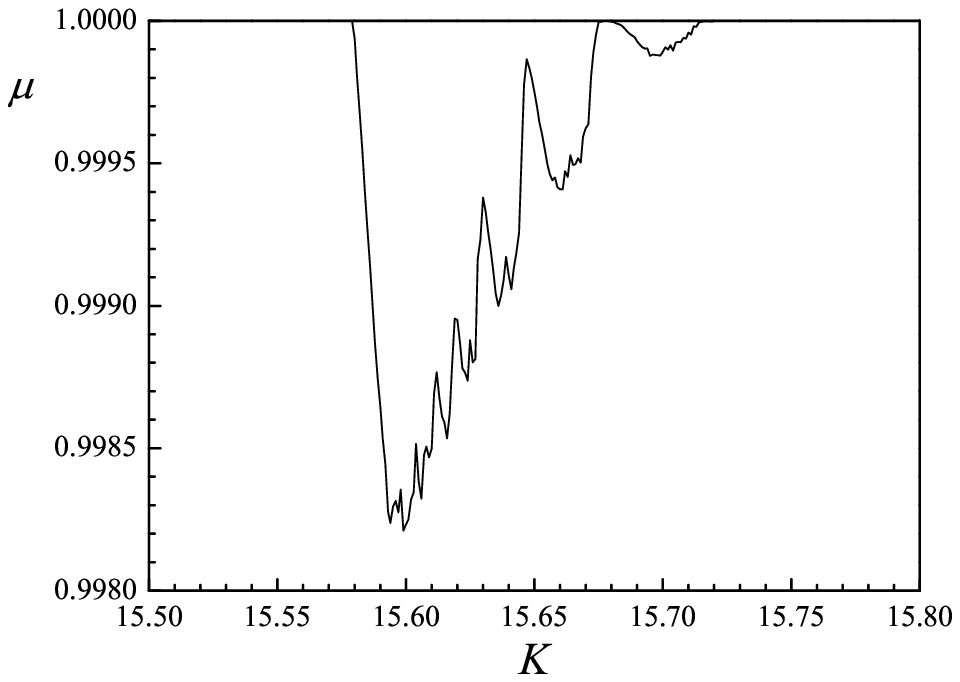} &
b)~\includegraphics[width=7cm]{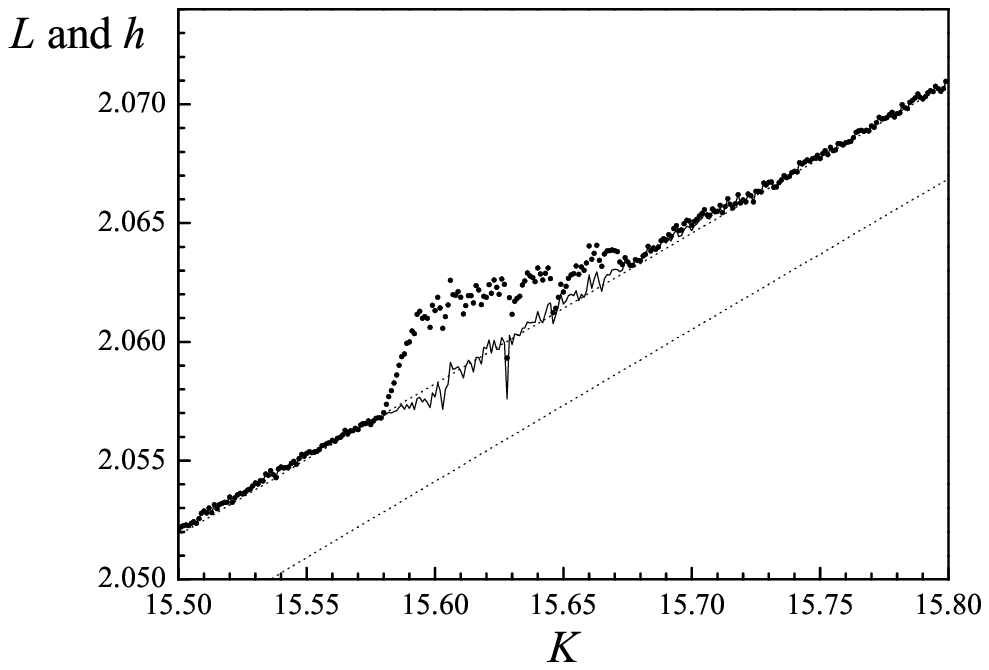}
\end{tabular}
\end{center}
\caption{The $\mu(K)$ dependence for a pattern at $K \approx 15.6$
due to a 4-periodic solution~(a); $L(K)$ and $h(K)$, drawn by bold
dots and solid curve respectively, for the same pattern~(b). The
thin dotted lines are the same as the thin solid lines in Fig.~2.}
\label{fig3}
\end{figure}

\begin{figure}
\begin{center}
\begin{tabular}{ll}
a)~\includegraphics[width=6.8cm]{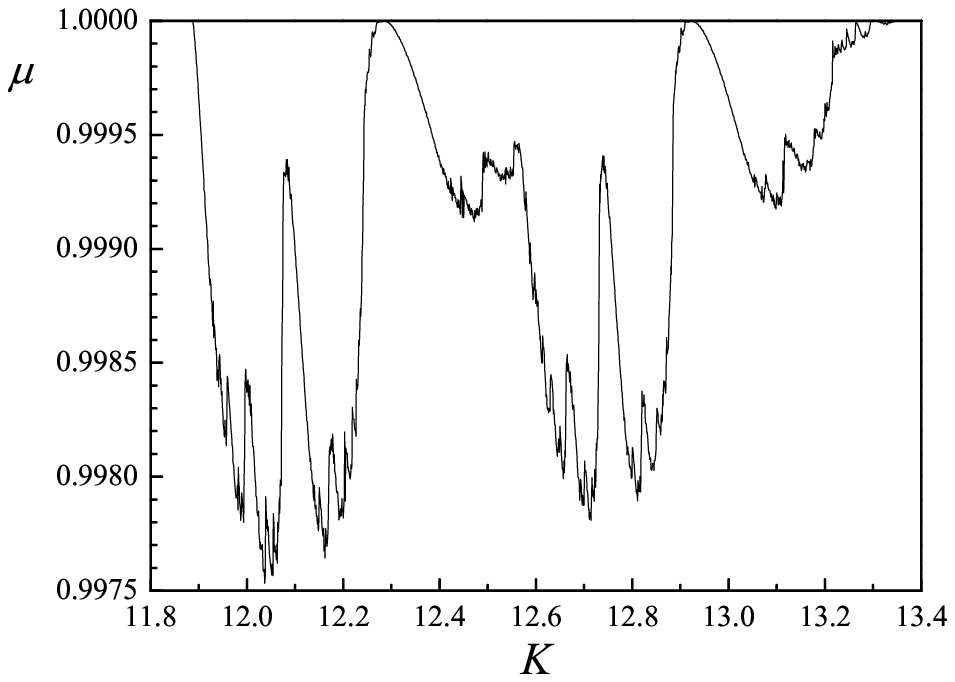} &
b)~\includegraphics[width=7cm]{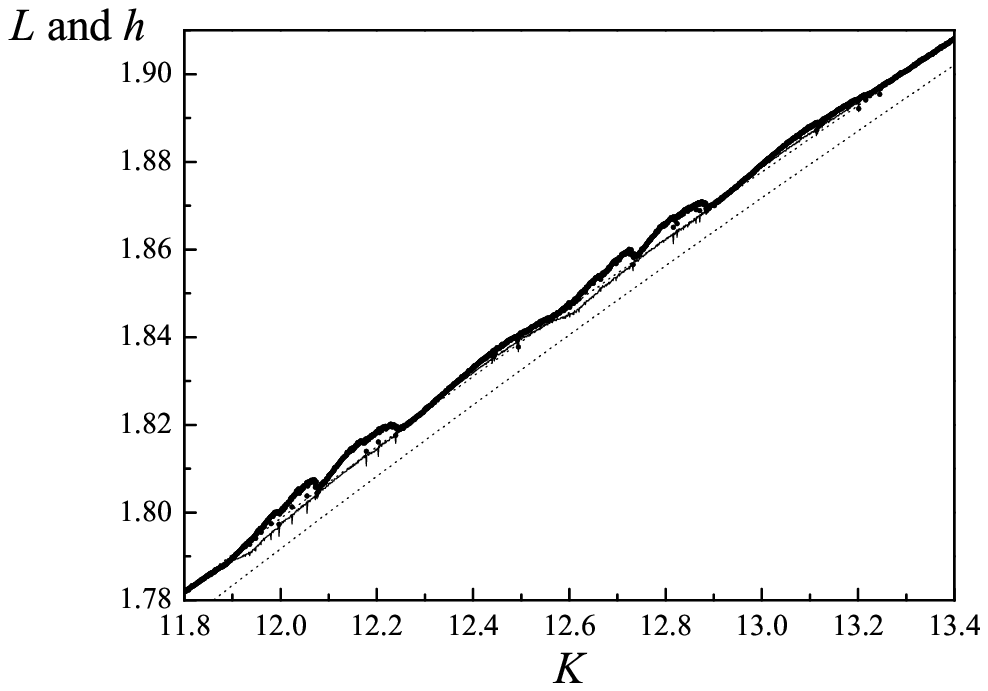}
\end{tabular}
\end{center}
\caption{The $\mu(K)$ dependence for a pattern at $K \approx 12
\div 13$ due to an accelerator mode~(a); $L(K)$ and $h(K)$ for the
same pattern~(b). The thin dotted lines are the same as the thin
solid lines in Fig.~2.}
\label{fig4}
\end{figure}

\end{document}